\documentclass[twocolumn, pra,showpacs,superscriptaddress]{revtex4}
\usepackage{graphicx}
\usepackage{dcolumn}
\usepackage{bm}
\usepackage{amsmath}

\begin{document}

\title{One-dimensional hard-core anyon gas in a harmonic trap at finite temperature}
\author{Yajiang Hao}
\email{haoyj@ustb.edu.cn}
\affiliation{Department of Physics, University of Science and Technology Beijing, Beijing 100083, China}
\author{Yafei Song}
\affiliation{Department of Physics, University of Science and Technology Beijing, Beijing 100083, China}
\date{\today }

\begin{abstract}
We investigate the strongly interacting hard-core anyon gases in a one dimensional harmonic potential at finite temperature by extending thermal Bose-Fermi mapping method to thermal anyon-ferimon mapping method. With thermal anyon-fermion mapping method we obtain the reduced one-body density matrix and therefore the momentum distribution for different statistical parameters and temperatures. At low temperature hard-core anyon gases exhibit the similar properties as those of ground state, which interpolate between Bose-like and Fermi-like continuously with the evolution of statistical properties. At high temperature hard-core anyon gases of different statistical properties display the same reduced one-body density matrix and momentum distribution as those of spin-polarized fermions. The Tan's contact of hard-core anyon gas at finite temperature is also evaluated, which take the simple relation with that of Tonks-Girardeau gas $C_b$ as $C=\frac12(1-cos\chi\pi)C_b$.
\end{abstract}

\maketitle

\section{introduction}

Basing on the symmetry under exchange satisfied by identical particles, quantum particles might be bosons or fermions. In one and two dimension there exists anyons \cite{anyons} satisfying fractional statistics that interpolate between bosons and fermions continuously \cite{Wilczek}. Fractional statistics not only play important roles but also become an important concept in condensed matter physics\cite{Wilczek,Halperin,Laughlin,Camino,YSWu,ZNCHa}. For the topological protection of quantum coherence of quantum system of fractional statistics \cite{RMP2008,Kitaev} and the potential application in quantum information science, search and realization of quantum system satisfying fractional statistics have become more and more important \cite{RMP2008,Mong}. For the high controllability and tunability quantum gas is a key candidate of creating anyons. Besides the rotated Bose-Einstein condensates \cite{PZoller}, cold atoms in optical lattices have become the popular platform \cite{DuanLM,OL}. The critical idea is the manipulation of transition rate as atom tunnel between lattice sites. It has been proposed that the one-dimensional (1D ) anyon gas can be prepared with the Raman-assisted hopping \cite{NatureComm,LSantos} or the lattice-shaking-induced tunneling \cite{Strater}.

The 1D quantum gas \cite{1D} can be realized by confining cold atoms in highly anisotropic trap and in optical lattices \cite{Ketterler,Paredes,Toshiya}. Both the arrays of 1D quantum gases and a single 1D quantum gas of strong interaction can be performed experimentally \cite{1DFiniteT}. The latter allow us to study the temperature dependent properties \cite{Olf} and the interplay effect between interaction and temperature. Since the realization of 1D quantum gases, those previous "toy models" in textbooks help us understand important physics by parameter-free comparison of theoretical prediction with measurement. With the Feshbach resonance technique and confinement-induced resonance technique we can realize the 1D quantum gas in the full interacting regime from weak to infinite strong interaction \cite{Paredes,Toshiya,Jacqmin,LSantos}. The Feshbach resonance technique can also be utilized to tune the interaction strength of 1D anyon gas \cite{LSantos}.

Theoretically although 1D anyon gas originated in condensed matter physics \cite{Haldane,WangZD}, it is found that 1D Bose gas with double $\delta$ function interaction is equivalent to the $\delta$-anyon gas \cite{Kundu99,Girardeau06}. The $\delta$-anyon gas attracted many theoretical interests in its ground state properties including  the exact solution \cite{Kundu99,Girardeau06,Batchelor}, correlation function \cite{Patu07,Patu08,Calabrese}, entanglement properties \cite{Cabra,HLGuo}, momentum distribution and the reduced one-body density matrix (ROBDM) \cite{Cabra,anyonTG,HaoPRA78,HaoPRA79,Campo,HaoPRA2012}. In addition relaxation dynamics \cite{MRigol} and quantum walks \cite{YZhang} were paid attentions to. It has been shown that the properties dependent on wavefunction rather than its modulus exhibit behaviours dependent statistical properties. For example, ROBDM become complex rather than real and momentum distributions become asymmetric rather than symmetric for anyon gases \cite{Cabra,anyonTG,HaoPRA78,HaoPRA79,Campo,HaoPRA2012,HaoPRA2016}. So far much investigations focus on anyon gas at zero temperature, and the temperature effect was not received much attentions and only the formal solutions of hard-core anyons at finite temperature were obtained \cite{Patu07,Patu08} although in experiments temperature is an important variable. In thermodynamic limit the solution of homogeneous anyon gas can also be evaluated with Bethe ansatz.

In the present paper we will follow the procedure in Ref. \cite{Lenard} and \cite{Minguzzi} to formulate the ROBDM of hard-core anyon gas at finite temperature with the determinant of one-body density matrix of spin-polarized fermions. For the first time we obtain its momentum distribution at finite temperature and extract temperature and statistical property dependence of the high-momentum tail. The Tan's contact coefficients will also be evaluated for different temperature and statistical properties. In the strong interaction regime the universal property is an important issue and several groups have investigated strongly interacting fermions \cite{OHara,Bourdel,Ku}. It was shown that the momentum distribution of Fermi gas has a tail falling off like $k^{-4}$ \cite{TanMom,Braaten,Stewart} and other properties including thermodynamics are also related the coefficient of the tail \cite{TanEnergy,TanVirial,SZhang}. The power-law decay of 1D Bose gas also satisfy $k^{-4}$ at large momentum \cite{Olshanii2003,Minguzzi,MinguzziPLA} and the contact coefficient has been evaluated. The present work will focus on the contact coefficient of 1D hard-core anyon gas whose momentum distribution is asymmetric about zero momentum.

The paper is organized as follows. In Sec. II, we briefly review the method to obtain the ROBDM of 1D hard-core anyon gas. In Sec. III, we present
the ROBDM and momentum distributions. The Tan's contact for different statistical parameters at finite temperature is investigated in Sec. IV. The summary is given in Sec. V.


\section{model and method}

We consider $N$ anyons of mass $m$ with the infinite repulsive contact interaction trapped in a harmonic potential
\begin{equation}
V_{ext}=m\omega ^2x^{2}/2.
\end{equation}

In the infinitely strong repulsion limit the many body wavefunction $\psi^A(x_1,x_2,...,x_N)$ satisfies the eigen equation
\begin{equation}
H\psi^A(x_1,x_2,...,x_N)=E\psi^A(x_1,x_2,...,x_N)		  \label{EigenEq}
\end{equation}
with $H=\sum_{i=1}^N\left [-\frac{\hbar^2}{2m}\frac{\partial^2}{\partial x_i^2}+\frac12m\omega^2x_i^2\right ]$ and the constraint condition
\begin{equation}
\psi^A(x_1,...,x_i,...,x_j,...,x_N)=0		\label{BC}
\end{equation}
if $x_i=x_j$ ($1\leq i < j \leq N$). Since the constraint condition can be satisfied by the wavefunction of spin-polarized fermions, all eigen wavefunctions $\psi^A(x_1,x_2,...,x_N)$ of eigen equation Eq.(\ref{EigenEq}) of anyons can be obtained with the wavefunction of spin-polarized fermions $\psi^{F}(x_{1},x_{2},...,x_{N})$ with the anyon-fermion mapping method \cite{Girardeau06}
\begin{equation}
\psi ^{A}(x_{1},\cdots ,x_{N})=\mathcal{A}(x_{1},\cdots ,x_{N})\psi^{F}\left( x_{1},x_{2},\cdots ,x_{N}\right) .
\end{equation}%
Here the anyonic mapping function is formulated as
\begin{equation}
\mathcal{A}(x_{1},\cdots ,x_{N})=\prod_{1\leq j<k\leq N}\exp [-\frac{i\chi \pi }{2}\epsilon (x_{j}-x_{k})]
\end{equation}%
with $0\leq \chi \leq 1$ being the statistical parameter. 1 corresponds to the hard-core bosons and 0 corresponds to noninteracting fermions. The sign function $\epsilon(x)$ gives $-1$, 0, and 1 depending on whether $x$ is negative, zero, or positive.

The wavefunction of $N$ polarized fermions $\psi ^F\left(x_{1},x_{2},\cdots,x_{N}\right) $ can be constructed by the one-particle wavefunction as
\begin{equation}
\psi ^{F}\left( x_{1},x_{2},\cdots ,x_{N}\right) =\left( 1/\sqrt{N!}\right)\det_{j,k=1}^{N}\phi _{\nu _j}\left( x_{k}\right) .
\end{equation}%
Here $\phi _{\nu _j}(x)$ is the $\nu _j$th eigen wavefunction of one particle in a harmonic trap $\phi_{\nu _j}(x)=(\sqrt{\pi}2^{\nu _j}\nu_j!/a_0)^{-\frac{1}{2}}e^{-a_0^2x^{2}/2}H_{\nu_j}(a_0x)$ with $H_{\nu_j}(x)$ being Hermite polynomial. $a_0=\sqrt{\hbar/m\omega}$ is the characteristic length of harmonic oscillator. The eigen wavefunctions correspond to one of the sets $\alpha=\{\nu_1,\nu_2,...,\nu_N\}$, where $\nu_j$ ($j=1,\cdots ,N$) are unequal positive integers. With the sets $\alpha$ all ground state and excited states satisfying the eigen equation $H\psi^A_{N,\alpha}(x_1,x_2,...,x_N)=E_{N,\alpha}\psi^A_{N,\alpha}(x_1,x_2,...,x_N)$ can be obtained with eigen energy $E_{N,\alpha}=\sum_{i=1}^N(\nu_i+1/2)\hbar \omega$.

At finite temperature $T$ the ROBDM of hard-core anyon gas in the grand-canonical ensemble is
\begin{multline}\label{eq6}
\rho_{1A}(x,y) =\sum_{N,\alpha}P_{N,\alpha}N\int_{-\infty}^{\infty}dx_2 \cdots dx_N \\
\times \psi^A_{N,\alpha}(x,x_2\cdots x_N)\psi^{A*}_{N,\alpha}(y,x_2\cdots x_N)
\end{multline}
The thermal distribution function reads $P_{N,\alpha}=Z^{-1}e^{-(E_{N,\alpha}-\mu N)/k_BT}$ with partition function $Z=\sum_{N,\alpha}e^{-(E_{N,\alpha}-\mu N)/k_BT}$ and chemical potential $\mu$.

Inserting the eigen functions of anyons Eq. (4) into the above integral and for each variable rewriting the integral $\int_{-\infty}^{\infty} dx_i e^{-\frac{i\chi\pi}{2}\left[\epsilon\left(x-x_i\right)-\epsilon\left(y-x_i\right)\right]}f =\int_{-\infty}^{\infty} dx_if-\epsilon(y-x)(1-e^{i\chi\pi\epsilon(y-x)})\int_x^ydx_i f$ we have
\begin{multline}\label{eq10}
\rho_{1A}(x,y) =\sum_{N,\alpha}P_{N,\alpha}N\sum_{j=0}^{N-1}
\begin{pmatrix}
N-1\\
j
\end{pmatrix}
(-1)^j \\
\times \left[\epsilon(y-x)(1-e^{i\chi\pi\epsilon(y-x)})\right]^j\int_x^y dx_2\cdots dx_{j+1}\\
\times \int_{-\infty}^{\infty} dx_{j+2}\cdots dx_N \psi^A_{N,\alpha}(x,x_2\cdots x_N)\psi^{A*}_{N,\alpha}(y,x_2\cdots x_N).
\end{multline}
Following the procedure in Ref. \cite{Lenard} and \cite{Minguzzi}, the ROBDM can be reformulated in terms of the fermionic $j$-body density matrix
\begin{eqnarray}\label{eq10}
&&\rho_{1A}(x,y) =\sum_{j=0}^{\infty}\left[-\epsilon(y-x)(1-e^{i\chi\pi\epsilon(y-x)})\right]^j/j!  \\
&&\times\int_x^y dx_2\cdots dx_{j+1}\rho_{j+1,F}(x,x_2,\cdots,x_N;y,x_2,\cdots,x_N),  \nonumber
\end{eqnarray}
where the fermionic $j$-body density matrix
\begin{multline}\label{eq13}
\rho_{jF}\left(x_1,\cdots,x_j,x'_1,\cdots,x'_j\right)=det\left[\rho_{1F}(x_i,x'_l)\right]_{i,l=1\cdots j}
\end{multline}
with the fermionic one body density matrix $\rho_{1F}(x,y)=\sum_{j=1}^Nf_{\nu _j}\phi_{\nu_j}(x)\phi^*_{\nu _j}(y)$ and Fermi-Dirac distribution $f_{\nu}=1/[e^{((\nu+1/2)\hbar\omega-\mu )/k_BT}+1]$.

Utilizing the properties of determinant the multiple integral in above formula can be reduced into the product of single-variable integral and the ROBDM of anyons will be reformulated as
\begin{multline}\label{eq15}
\rho_{1A}\left(x,y\right)=\sum_{j=0}^{\infty}\frac{[-\epsilon(y-x)(1-e^{i\chi\pi\epsilon(y-x)})]^j}{j!}\rho_{1A}^{(j)}\left(x,y\right),
\end{multline}
where
\begin{multline}\label{eq14}
\rho_{1A}^{(j)}\left(x,y\right)=\sum_{\nu_1\cdots \nu_{j+1}}f_{\nu_1}\cdots f_{\nu_{j+1}}\sum_{P\in S{j+1}}(-1)^P \\
\phi_{\nu_1}(x)\phi_{\nu_{P(1)}}(y)\prod_{l=2}^{j+1}\int_x^ydx_l\phi_{\nu_l}(x_l)\phi^*_{\nu_{P(l)}}(x_l).
\end{multline}
Its diagonal part $\rho(x)=\rho_{1A}(x,x)$ is density distribution of anyons, which is independent on the statistical parameter. In the Bose limit $\chi =1.0$, the above formula reduces to the same result as that in Ref. \cite{Minguzzi}. The momentum distribution of anyons can be obtained by the Fourier transform of ROBDM
\begin{equation}
n(k)=(2\pi)^{-1}\int_{-\infty}^{\infty}dxdy\rho_{1A}\left( x,y\right) e^{-ik(x-y)}.  \label{MD}
\end{equation}

In the following sections we will display the temperature effect on the ROBDM and momentum distribution of 1D hard-core anyons. For simplification the natural unit will be used and the present notation will be preserved.

\section{ROBDM and momentum distribution of anyon gas at finite temperature}

In this section, we evaluate the ROBDM and momentum distribution of 1D strongly interacting anyon gases of $N$ anyons in a harmonic trap for different statistical properties at finite temperature.

\begin{figure}[tbp]
\includegraphics[width=3.0in]{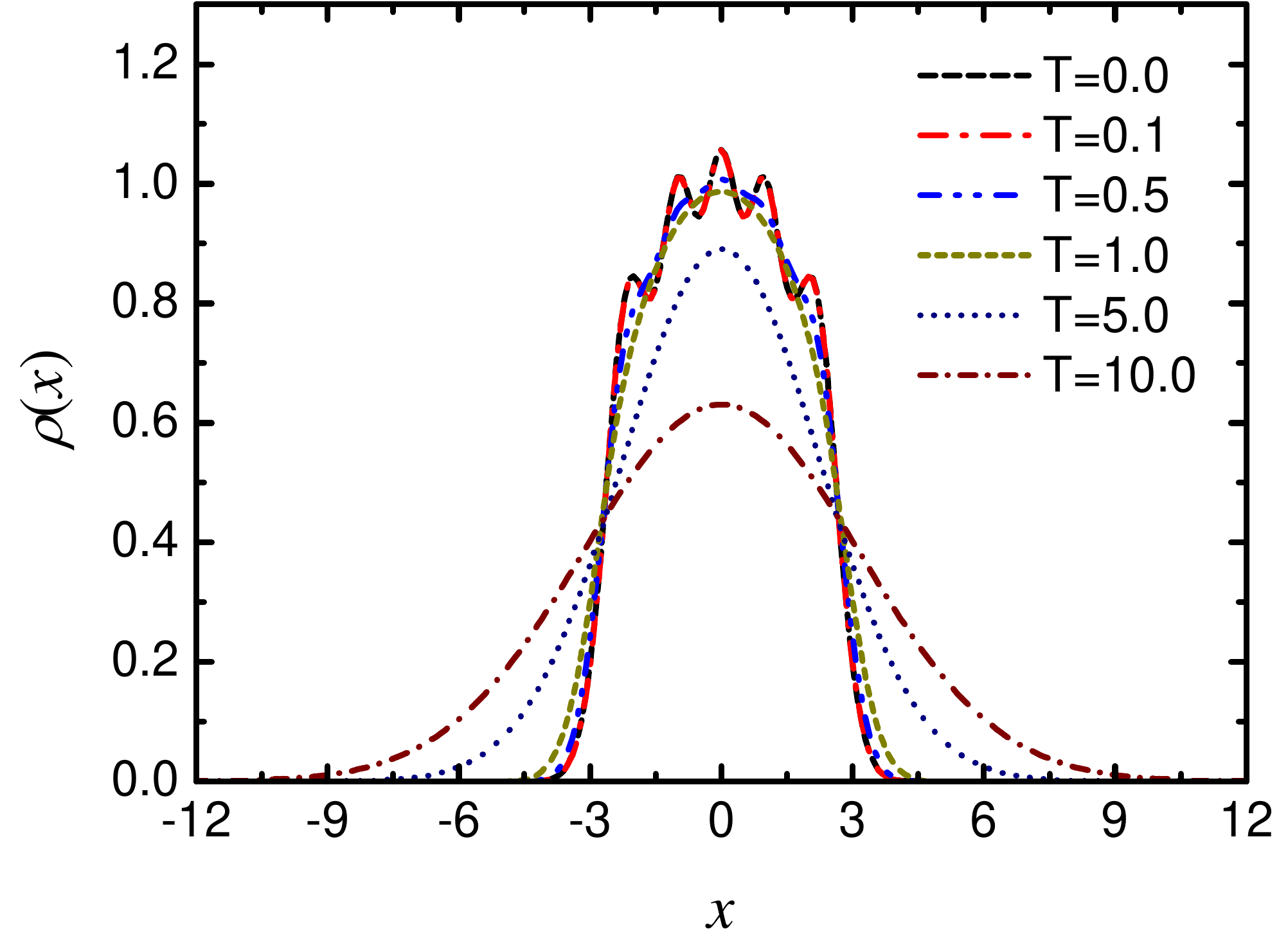}
\caption{The density distribution of hard-core anyon gas of $N$=5 at finite temperature. Unit of $x$: $a_0=\sqrt{\hbar/m\omega}$.}
\end{figure}

In Fig. 1 we display the diagonal part of ROBDM, i.e., the density distributions $\rho(x)$, of hard-core anyon gases with $N=5$. According to Eq. (5), the density distributions are not related to the statistical properties. It is shown that at low temperature ($T=0.1$) the temperature effect is not displayed, and the density profile show the same shell structure of $N$ peaks as ground state density profiles, which are also same as those of Tonks gas and spin-polarized Fermi gas at zero temperature. With the increase of temperature the shell structure disappears and the anyons still stay in the central region of the harmonic trap with large probability ($T$=0.5 and 1.0). At high temperature ($T$=5.0 and 10.0) anyons occupy in wider region because the higher kinetic energy increase the probability that anyons distribute in the region of high potential energy. The high temperature also induce the population in the high energy levels of single particle such that the density profiles behave similar to the Gaussian distribution.

\begin{figure}[tbp]
\includegraphics[width=3.2in]{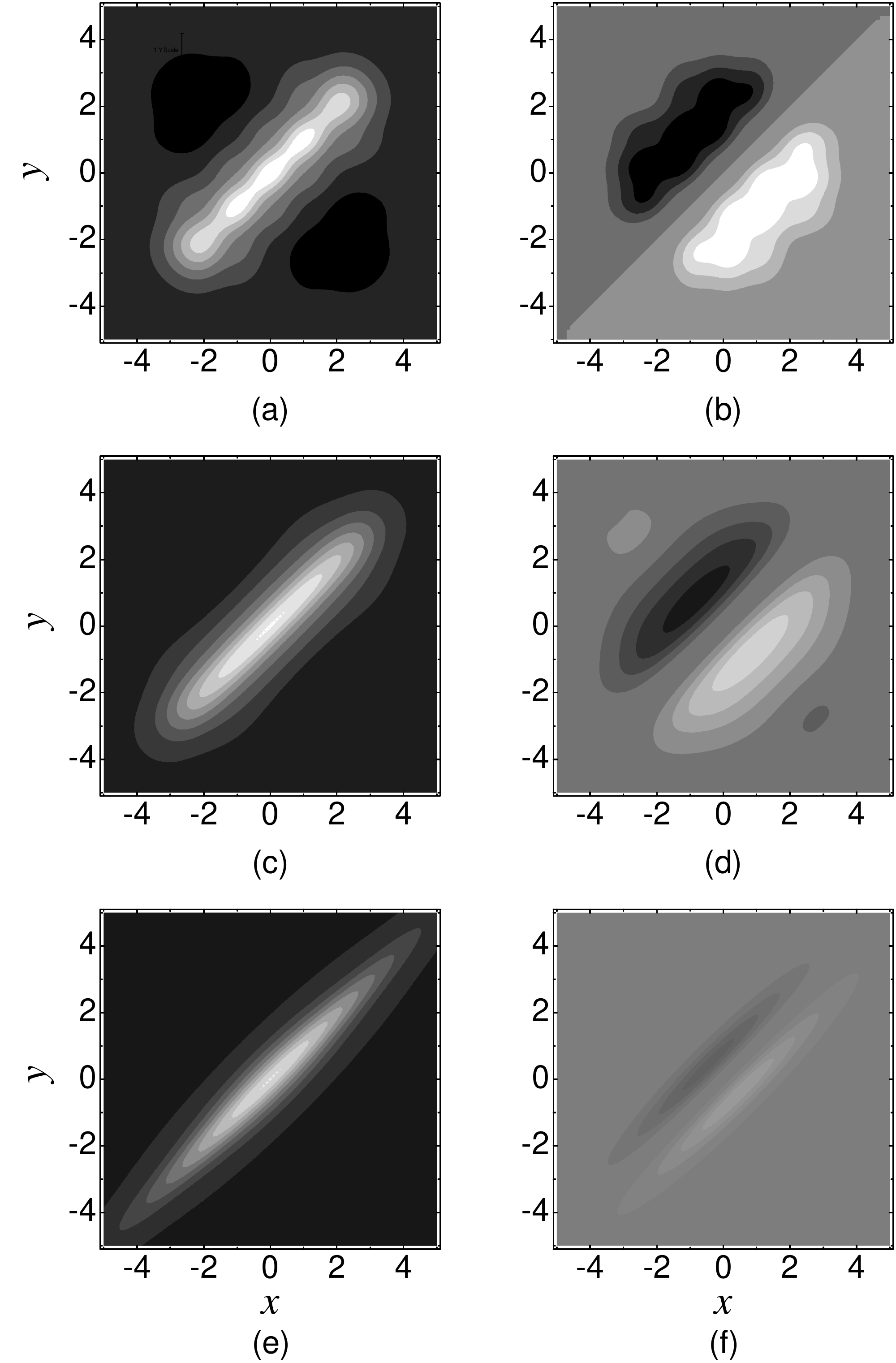}
\caption{The ROBDM of hard-core anyon gas of $N$=5 with statistical parameter $\chi=0.75$. Real part of ROBDM: (a) $T=0.1$, (c) $T=1.0$, and (e) $T=5.0$; Imaginary part of ROBDM:  (b) $T=0.1$, (c) $T=1.0$, and (f) $T=5.0$. Unit of $x$ and $y$: $a_0=\sqrt{\hbar/m\omega}$; Unit of $T$: $\hbar\omega/k_B$.}
\end{figure}

It has been shown that the ROBDM of ground states of anyons is complex rather than real \cite{HaoPRA78,HaoPRA79,HaoPRA2012,HaoPRA2016}. The ROBDM of hard-core anyon gas of $N=5$ with statistical parameter $\chi=0.75$ are displayed in Fig. 2 at finite temperature. The ROBDM are still complex and we display the real part (left column) and imaginary part (right column), respectively. The real part are symmetric matrix, while the imaginary part are antisymmetric, which will result in the asymmetric momentum distribution. At low temperature, the real part are diagonal dominant but the off-diagonal matrix elements are not negligibly small, which embody the long-range order of anyon gases of $\chi=0.75$. With the increase of temperature, not only the off-diagonal matrix elements of real part become negligible but also the full imaginary part approximate to zero. At high temperature the ROBDM of anyon gases exhibit the same properties as those of Tonks gas and spin-polarized fermions \cite{HaoIJMPB2016}.

\begin{figure}[tbp]
\includegraphics[width=3.0in]{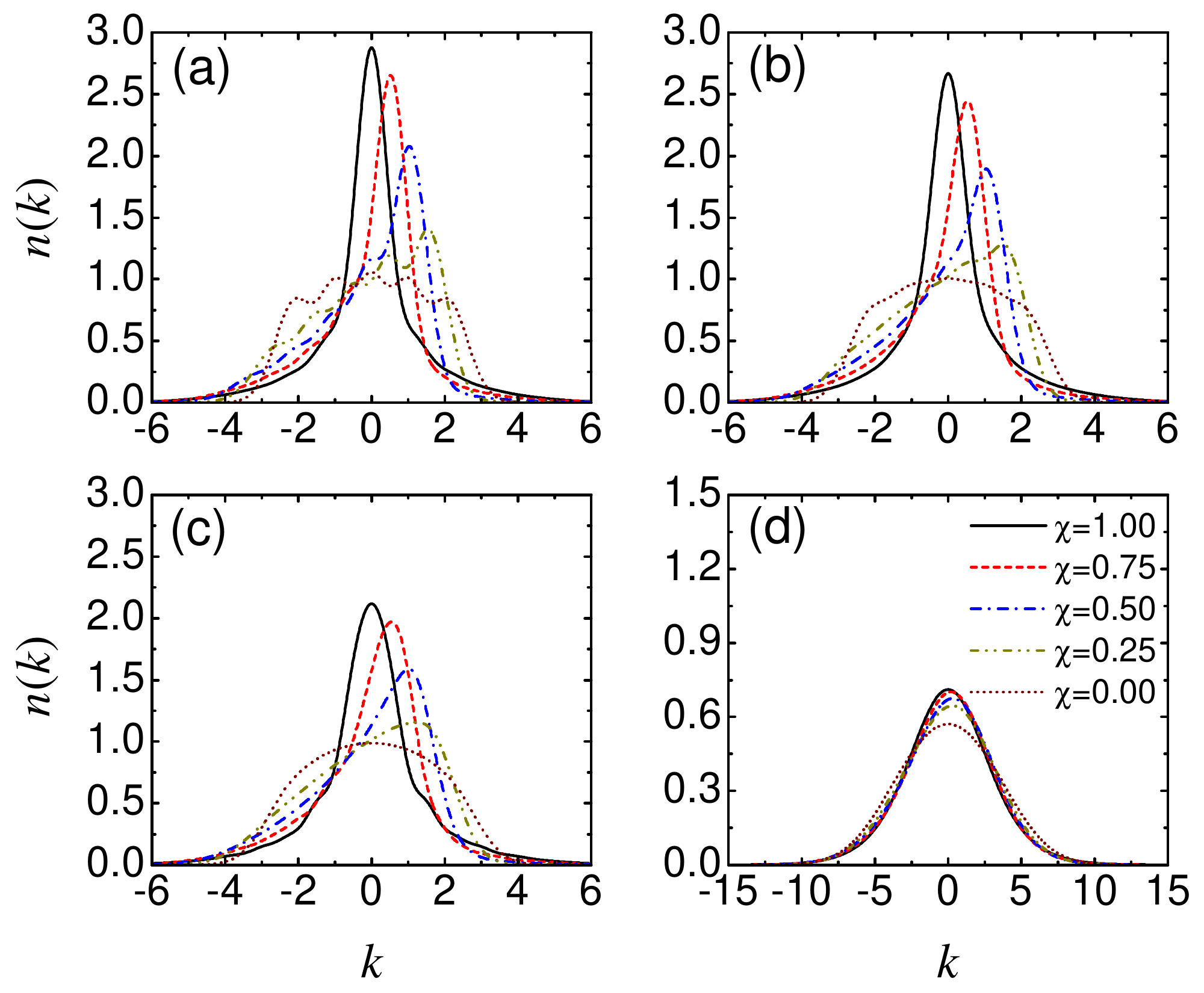}
\caption{Momentum distribution of hard-core anyon gas of $N$=5 at different temperatures. (a) $T=0.1$; (b) $T=0.5$; (c) $T=1.0$; (d) $T=10.0$. Unit of $k$: $a_0^{-1}$; Unit of $T$: $\hbar\omega/k_B$.}
\end{figure}

We displayed the momentum distribution of hard-core anyon gas of $N=5$ at finite temperature in Fig. 3. It is shown that momentum distributions are asymmetric about the zero momentum at finite temperature except those in the Bose limit ($\chi=1.0$) and Fermi limit ($\chi=0.0$). At low temperature ($T$=0.1) Bosons exhibit single sharp peak momentum profile and distribute in the zero momentum region with great probability, while Fermions display shell structure of $N$ peaks. With the decrease of statistical parameter, the peak of momentum distributions of anyons shift away from zero momentum and then shift back in the Fermi limit. This is same as the case of ground state at zero temperature. As temperature increases ($T$=0.5 and 1.0), the oscillation of momentum profiles become weak and anyons exhibit smooth momentum distribution even in the Fermi limit. At the same time, anyons distribute in high momentum regions with larger probability and the peak height of momentum profiles decrease. Another temperature effect on momentum distribution is that the asymmetry become obscure, which is exhibited at high temperature ($T$=10.0). In this situation it is hard to distinguish the statistical properties of anyons by the momentum profile and anyons with different statistical parameters behave similar momentum distributions.


\section{Tan's contact of hard core anyon gas at finite temperature}

The momentum distribution of 1D Bose gas decays as the power-law $Ck^{-4}$ at large momentum \cite{MinguzziPLA,Olshanii2003} and the temperature dependence of Tan's contact $C$ was shown in Ref. \cite{Minguzzi}. It is interesting to investigate the temperature and statistical properties dependence of the universal power-law of hard-core anyon gas at high momenta.

Following the procedure in Ref. \cite{Minguzzi}, the high-momentum tails are related to the short-distance behaviour of one body density matrix and the main contribution comes from the $j=1$ term. Expanding $\rho_{1A}(x,y)$ at $R=(x+y)/2$ and retaining the lower-order term of $|x-y|$, we have
\begin{equation}
\rho _{1A}\left( x,y\right) \propto \frac{\left\vert x-y\right\vert ^{3}}{6}(1-e^{i\chi \pi \epsilon (y-x)})f\left( R\right),
\end{equation}
where
\begin{equation*}
f\left( R\right) = n(R)\sum_{\nu }f_{\nu }\left\vert \phi _{\nu
}^{\prime }\left( R\right) \right\vert ^{2}-\left\vert \sum_{\nu }f_{\nu
}\phi _{\nu }\left( R\right) \phi _{\nu }^{\ast \prime }\left( R\right)
\right\vert
\end{equation*}
with $n\left( R\right)=\sum_{\nu} f_{\nu }\phi _{\nu }\left( R\right) \phi _{\nu }^{\ast}\left( R\right) $. Using the asymptotic expansion of Fourier transformations in the limit of large $k$ \cite{FourierAnalysis}
\begin{eqnarray*}
&&\lim_{k\rightarrow \infty }\int_{-\infty }^{\infty }dxe^{-2\pi ikx} \left\vert x\right\vert ^{\alpha }dx \\
&=&2\cos \frac{\left( \alpha
+1\right) \pi }{2}\alpha !\left( 2\pi \left\vert k\right\vert \right)
^{-\alpha -1},
\end{eqnarray*}%
and
\begin{eqnarray*}
&&\lim_{k\rightarrow \infty }\int_{-\infty }^{\infty }dxe^{-2\pi ikx} \left\vert x\right\vert ^{\alpha }\epsilon \left(x\right) dx  \\
&=&-2i\sin \frac{\left( \alpha
+1\right) \pi }{2}\alpha !\left( 2\pi \left\vert k\right\vert \right)
^{-\alpha -1}\epsilon \left( k\right) ,
\end{eqnarray*}%
we find that the momentum distribution of 1D hard-core anyon gas decay as $Ck^{-4}$ in the limit of $k \rightarrow \infty$ with the Tan's contact
\begin{equation}
C=\frac{1-cos(\chi \pi )}{\pi }\int_{-\infty }^{\infty }dRf\left( R\right).
\end{equation}
Compared with the Tan's contact of 1D Bose gas $C_b$, that of 1D hard-core anyon gas is $C=\frac12(1-cos(\chi \pi ))C_b$.

\begin{figure}[tbp]
\includegraphics[width=3.2in]{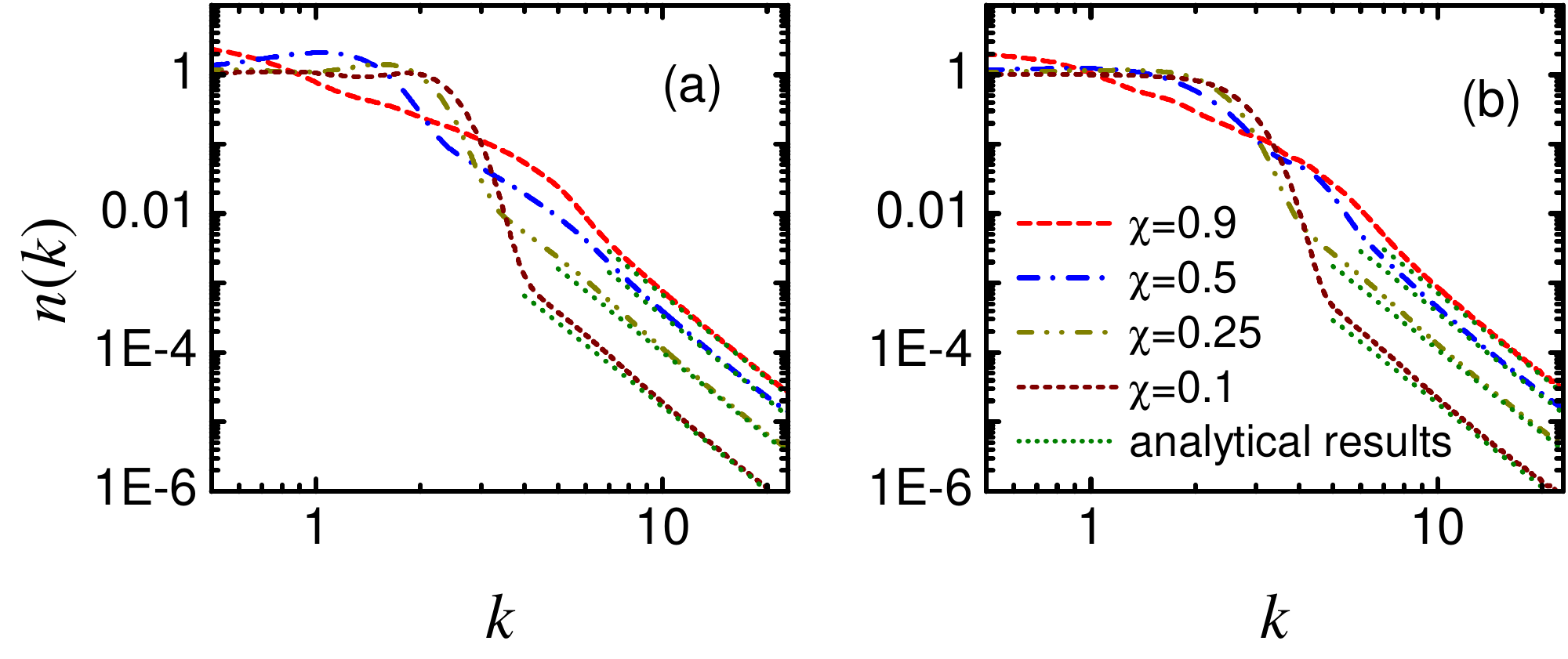}
\caption{The high momentum tail of momentum distribution for anyon gases of $N$=5 with different statistical parameter. (a) $T=0.1$; (b) $T=1.0$. The analytical result are plotted as dotted lines. The log-log coordinates are used. Unit of $k$: $a_0^{-1}$; Unit of $T$: $\hbar\omega/k_B$.}
\end{figure}

The high momentum tail of momentum distribution for anyon gases at finite temperature $T$=0.1 and 1.0 are plotted in Fig. 4, which are evaluated numerically by Eq. (11) and Eq. (13). As comparisons, the analytical result $Ck^{-4}$ with $C$ being determined by Eq. (15), are also plotted in dotted lines for different statistical parameters. It is shown that numerical result match well with the analytical results for all statistical parameters at finite temperatures. The momentum distribution of anyons decays as a power-law $Ck^{-4}$ at high momenta at finite temperature.

\begin{figure}[tbp]
\includegraphics[width=3.0in]{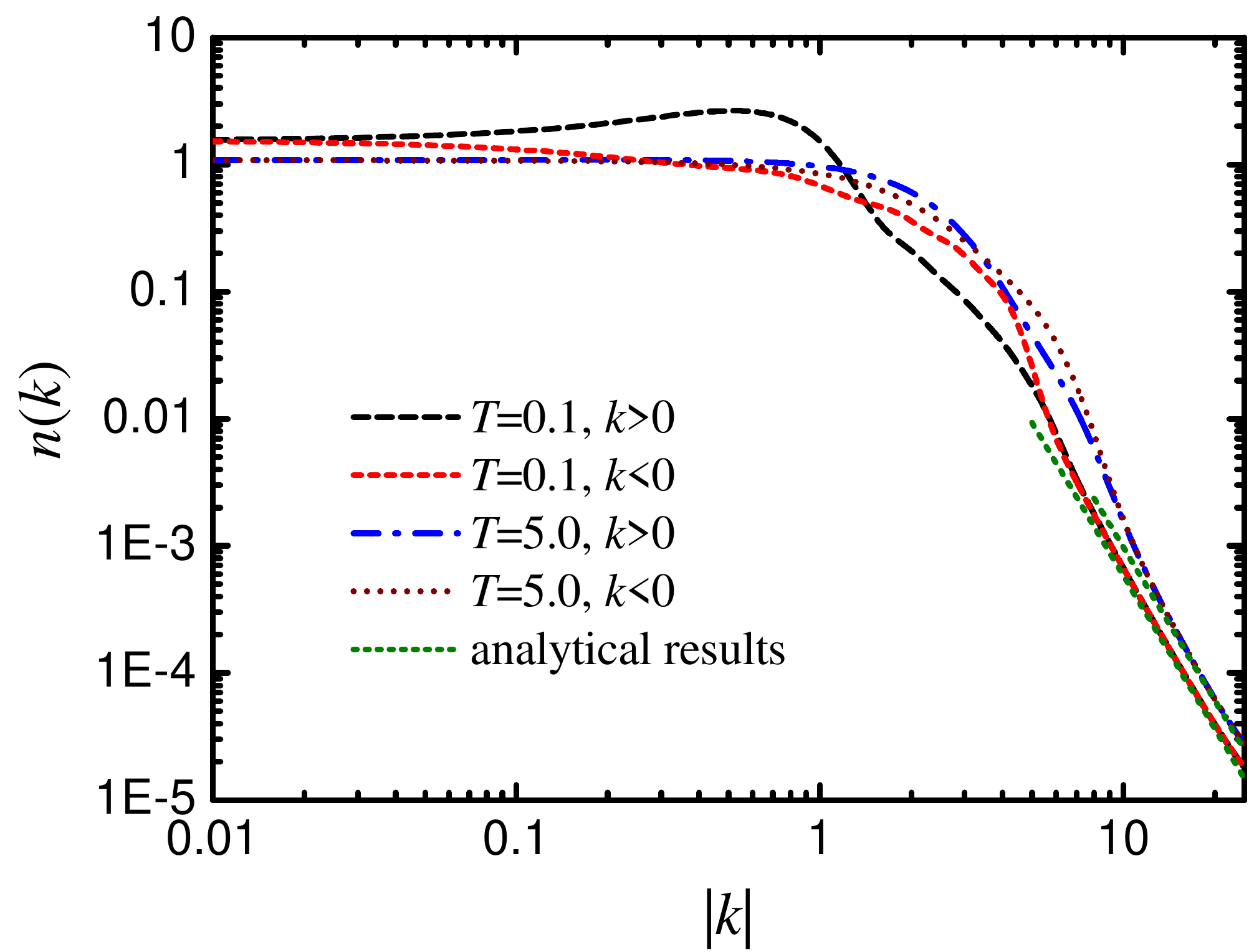}
\caption{The positive high momentum tail and negative high momentum tail of momentum distribution for anyon gas of $N$=5 with statistical parameter $\chi=0.75$ at finite temperature. The log-log coordinates are used. Unit of $k$: $a_0^{-1}$; Unit of $T$: $\hbar\omega/k_B$.}
\end{figure}

An interesting question is the asymmetry of momentum distribution of hard-core anyon gases. It has been shown that anyons exhibit different behaviours for positive momentum and negative momentum. Then the power-law behaviours dependence on the signature of momentum is worth to be investigated. The analytical evaluation manifest that the high momentum tail exhibit the same power-law behaviour for positive momentum and negative momentum although they are different for small momenta. In Fig. 5 we plot the positive high momentum tail and negative high momentum tail of hard-core anyon gas with $\chi=0.75$ at temperature $T$=0.1 and 5.0. It turns out that they match well with each other and match with analytical results at high momentum region both at low temperature ($T=0.1$) and at high temperature ($T=5.0$).

\begin{figure}[tbp]
\includegraphics[width=3.0in]{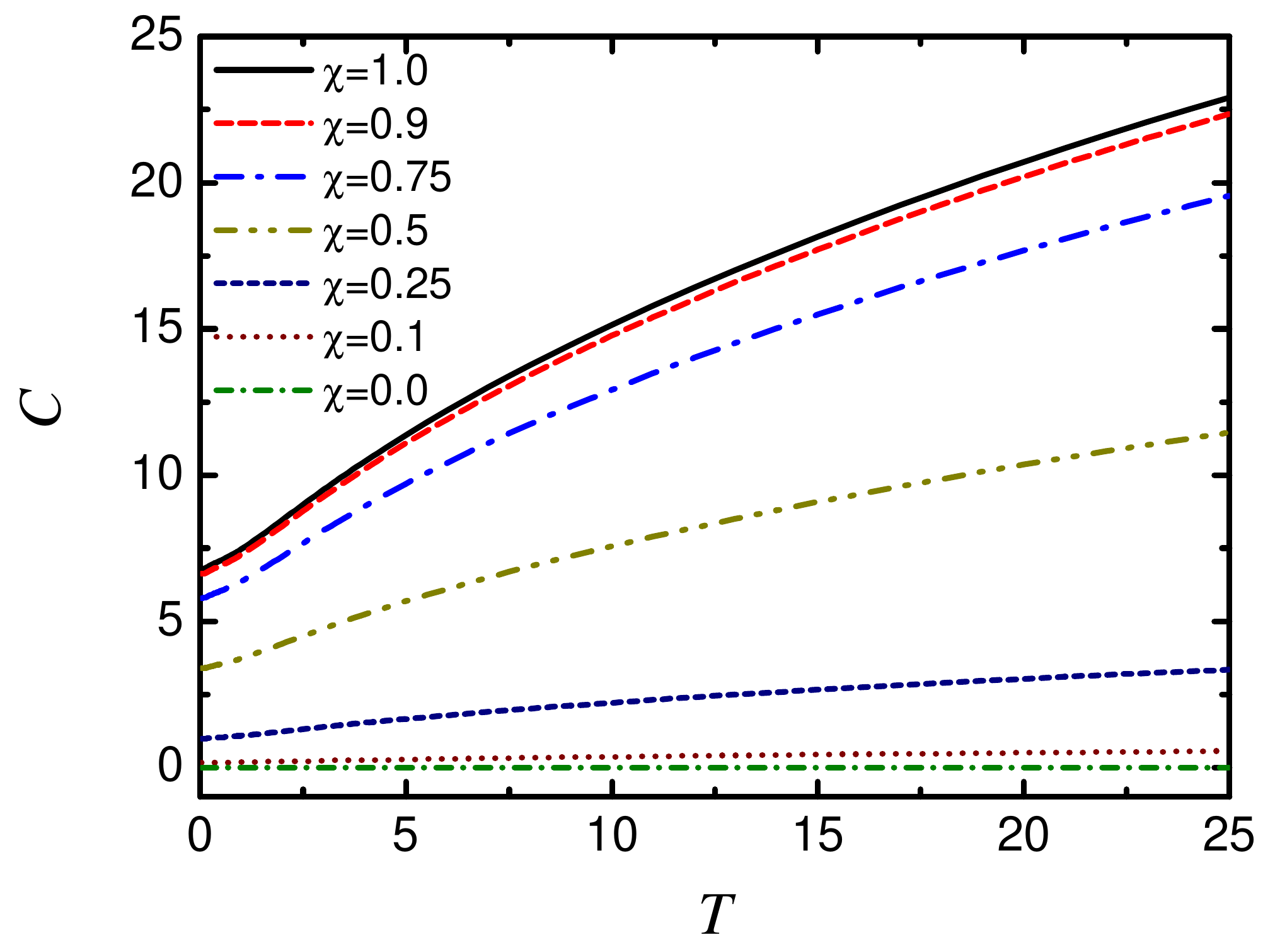}
\caption{ Tan's contact $C$ as a function of temperature $T$ for the anyon gas of $N$=5 with different statistical parameter. $C$ and $T$ are in units of $a_0^{-3}$ and $\hbar\omega/k_B$, respectively.}
\end{figure}

The temperature dependence of Tan's contact $C$ of hard-core anyon gases with different statistical properties are plotted in Fig. 6. In the Fermi limit ($\chi$=0.0) the Tan's contact is equal to zero in the full temperature regime, which is consistent to the Pauli exclusion principle satisfied by spin-polarized fermions whose interaction energy is always zero. As the statistical property deviates from the Fermi limit, the Tan's contact increase with the increasing temperature for hard-core anyon gases. At specific temperature the Tan's contact also increase with the statistical parameter. According to Eq. (15) we have $C \propto 1-cos(\chi\pi)$ at given temperature. The Tan's contact and therefore the interaction energy of hard-core Bosons has the maximum value.

\section{Summary}

In conclusion we investigated the 1D hard-core anyon gas confined in a harmonic trap at finite temperature with the thermal anyon-fermion mapping method. The thermal Bose-Fermi mapping method has been used to study the strongly interacting Tonks-Girardeau gas at finite temperature. It has been extended to investigate 1D strongly interacting anyon gas satisfying fractional statistics in the present work. By mapping eigen functions of spin-polarized fermions to eigen functions of hard-core anyons, we obtained the reduced one-body density matrix of hard-core anyons at finite temperature. Therefore the momentum distribution of hard-core anyons of statistical properties can be evaluated for different temperatures.

It was shown that similar to the ground-state properties of 1D hard-core anyons, at finite temperature the anyonic system also exhibit the asymmetric momentum distributions as the statistical properties deviate the Bose limit and Fermi limit. The asymmetry results from the properties of ROBDM, which is complex rather than real for anyon gases. At low temperature, the real part of ROBDM is diagonal dominant but the off-diagonal is not negligibly small, and its imaginary part is antisymmetry about $x=y$. With the increase of temperature, both the off-diagonal elements of real part and the full imaginary part become small. At high temperature, the imaginary part approx to zero and the ROBDM of hard-core anyon gas exhibit the same properties as that of spin-polarized fermions

Correspondingly, at low temperature region momentum distribution of anyon gas in Bose limit display the $\delta$-function-like single peak, and that in Fermi limit display shell structure of $N$ peaks. With the evolution of statistical property, momentum distribution of anyon gas evolve from Bose-like single peak structure into the Fermi-like shell structure, but the peak appears at finite momentum. At high temperature the asymmetry become weak and 1D hard-core anyon gas display almost symmetric momentum profiles. It is hard to distinguish the statistical property of anyon gas basing on the properties of momentum distribution in this case. The momentum distribution of 1D hard-core anyons with different statistical property exhibit the same profiles as that of spin-polarized fermions.

We also obtain the high-momentum tail of 1D hard-core anyon gas at finite temperature. Although the momentum distribution is asymmetric about zero momentum, the power-law decay at high momentum region is same for the positive momentum and negative momentum. This was proved both by the exact numerical calculation and by approximate analytical derivation. In high momentum region the momentum distribution of hard-core anyons decays as the power-law $Ck^{-4}$, which is same as that of 1D Bose gas. The Tan's contact depend on the statistical parameter and satisfy the relation $C=\frac12(1-cos\chi\pi)C_b$ with $C_b$ being the Tan's contact of 1D Bose gas.

\begin{acknowledgments}
This work was supported by NSF of China under Grants No. 11004007 and ``the
Fundamental Research Funds for the Central Universities."
\end{acknowledgments}


\begin{thebibliography}{99}
\bibitem{anyons} J. M. Leinaasand and J. Myrheim, Nuovo Cimento \textbf{37B}, 1 (1977); F. Wilczek, Phys. Rev. Lett. \textbf{49}, 957 (1982).

\bibitem{Wilczek} F. Wilczek, Fractional Statistics and Anyon Superconductivity, (World Scientific, Singapore 1990).

\bibitem{Halperin} B. I. Halperin, Phys. Rev. Lett. \textbf{52}, 1583 (1984).

\bibitem{Laughlin} R. B. Laughlin, Phys. Rev. Lett. \textbf{50}, 1395 (1983).

\bibitem{Camino} F. E. Camino, W. Zhou and V. J. Goldman, Phys. Rev. B \textbf{72}, 075342 (2005).

\bibitem{YSWu} Y.-S. Wu and Y. Yu, Phys. Rev. Lett. \textbf{75}, 890 (1995).

\bibitem{ZNCHa} Z. N. C. Ha, Phys. Rev. Lett. \textbf{73}, 1574 (1994); M. V. N. Murthy and R. Shankar, ibid. \textbf{73}, 3331 (1994); Z. N. C. Ha, Nucl. Phys. B \textbf{435}, 604 (1995).

\bibitem{RMP2008} C. Nayak, S. H. Simon, A. Stern, M. Freedman, and S. Das Sarma, Rev. Mod. Phys. \textbf{80}, 1083 (2008) .

\bibitem{Kitaev} A. Y. Kitaev, Ann. of Phys. \textbf{303}, 2 (2003).

\bibitem{Mong} R. S. K. Mong, D. J. Clarke, J. Alicea, N. H. Lindner, P. Fendley, C. Nayak, Y. Oreg, A. Stern, E. Berg, K. Shtengel et al., Phys. Rev. X \textbf{4}, 011036 (2014).

\bibitem{PZoller} B. Paredes, P. Fedichev, J. I. Cirac, and P. Zoller, Phys. Rev. Lett. \textbf{87}, 010402 (2001).

\bibitem{DuanLM} L.-M. Duan, E. Demler, and M. D. Lukin, Phys. Rev. Lett. \textbf{91}, 090402 (2003); A. Micheli, G. K. Brennen, and P. Zoller, Nature Phys. \textbf{2}, 341 (2006).

\bibitem{OL} C.-W. Zhang, V. W. Scarola, Sumanta Tewari, and S. Das Sarma, Proc. Natl. Acad. Sci. USA \textbf{104}, 18415 (2007); J.-K. Pachos, Ann. of Phys. \textbf{322}, 1254 (2007); M. Aguado, G. K. Brennen, F. Verstraete, and J. I. Cirac, Phys. Rev. Lett. \textbf{101}, 260501 (2008); L. Jiang, G. K. Brennen, A. V. Gorshkov, K. Hammerer, M. Hafezi, E. Demler, M. D. Lukin, and P. Zoller, Nat. Phys. \textbf{4}, 482 (2008).

\bibitem{NatureComm} T. Keilmann, S. Lanzmich, I. McCulloch, and M. Roncaglia, Nature Communications, \textbf{2}, 361 (2011).

\bibitem{LSantos} S. Greschner and L. Santos, Phys. Rev. Lett. \textbf{115}, 053002 (2015).

\bibitem{Strater} C. Str\"{a}ter, S. C. L. Srivastava, A. Eckardt, arXiv: 1602.08384v1.

\bibitem{1D} M. A. Cazalilla, R. Citro, T. Giamarchi, E. Orignac, and M. Rigol, Rev. Mod. Phys \textbf{83}, 1405 (2011); M. Olshanii, Phys. Rev. Lett. \textbf{81}, 938 (1998); D. S. Petrov, G. V. Shlyapnikov, and J. T. M. Walraven, Phys. Rev. Lett. \textbf{85}, 3745 (2000); V. Dunjko, V. Lorent, and M. Olshanii, Phys. Rev. Lett. \textbf{86}, 5413 (2001).

\bibitem{Ketterler} N. J. van Druten and W. Ketterle, Phys. Rev. Lett. \textbf{79}, 549 (1997).

\bibitem{Paredes} B. Paredes, A. Widera, V. Murg, O. Mandel, S. F\"{o}lling, I. Cirac, G. V. Shlyapnikov, T. W. H\"{a}nsch, and I. Bloch, Nature \textbf{429}, 277 (2004).

\bibitem{Toshiya} T. Kinoshita, T. Wenger and D. S. Weiss, Science \textbf{305}, 1125 (2004).

\bibitem{1DFiniteT} T. Jacqmin, J. Armijo, T. Berrada, K. V. Kheruntsyan, and I. Bouchoule, Phys. Rev. Lett. \textbf{106}, 230405 (2011).

\bibitem{Olf}  R. Olf, F. Fang, G. E. Marti, A. MacRae, and D. M. Stamper-Kurn, Nat. Phys. \textbf{11}, 720 (2016).

\bibitem{Jacqmin} T. Jacqmin, J. Armijo, T. Berrada, K. V. Kheruntsyan, and I. Bouchoule, Phys. Rev. Lett. \textbf{106}, 230405 (2011).


\bibitem{Haldane} F. D. M. Haldane, Phys. Rev. Lett. \textbf{67}, 937 (1991).

\bibitem{WangZD} J. X. Zhu and Z. D. Wang, Phys. Rev. A \textbf{53}, 600 (1996).

\bibitem{Kundu99} A. Kundu, Phys. Rev. Lett. \textbf{83}, 1275 (1999).

\bibitem{Girardeau06} M. D. Girardeau, Phys. Rev. Lett. \textbf{97}, 100402 (2006).

\bibitem{Batchelor} M. T. Batchelor, X. W. Guan, J. S. He, J. Stat. Mech.: Theor. Exp. P03007 (2007); M. T. Batchelor, X. W. Guan, Phys. Rev. B \textbf{74}, 195121
(2006); M. T. Batchelor, A. Foerster, X. W. Guan, J. Links, and H. Q. Zhou, J. Phys. A: Math. Theor. \textbf{41}, 465201 (2008).


\bibitem{Patu07} O. I. Patu, V. E. Korepin and D. V. Averin, J. Phys. A \textbf{40}, 14963 (2007).

\bibitem{Patu08} O. I. Patu, V. E. Korepin, and D. V. Averin, J. Phys. A \textbf{41}, 145006 (2008); J. Phys. A: Math. Theor. \textbf{41} 255205 (2008).

\bibitem{Calabrese} P. Calabrese and M. Mintchev, Phys. Rev. B \textbf{75}, 233104 (2007).

\bibitem{Cabra} R. Santachiara, R. F. Stauffer and D. Cabra, J. Stat. Mech.: Theor. Exp. L05003 (2007).

\bibitem{HLGuo} H. Guo, Y. Hao, and S. Chen, Phys. Rev. A \textbf{80}, 052332 (2009).

\bibitem{anyonTG} R. Santachiara and P. Calabrese, J. Stat. Mech.: Theor. Exp. P06005 (2008).

\bibitem{HaoPRA78} Y. Hao, Y. Zhang, and S. Chen, Phys. Rev. A \textbf{78}, 023631 (2008).

\bibitem{HaoPRA79} Y. Hao, Y. Zhang, and S. Chen, Phys. Rev. A \textbf{79}, 043633 (2009).

\bibitem{Campo} A. del Campo, Phys. Rev. A \textbf{78}, 045602 (2008).

\bibitem{HaoPRA2012} Y. Hao and S. Chen, Phys. Rev. A \textbf{86}, 043631 (2012).

\bibitem{MRigol} T. M. Wright, M. Rigol, M. J. Davis, and K. V. Kheruntsyan, Phys. Rev. Lett. \textbf{113}, 050601 (2014).

\bibitem{YZhang} L. M. Wang, L. Wang, and Y. Zhang, Phys. Rev. A \textbf{90}, 063618 (2014).

\bibitem{HaoPRA2016} Y. Hao, Phys. Rev. A \textbf{93}, 063627 (2016).

%
%

\bibitem{Lenard} A. Lenard, J. Math. Phys. 5, 930 (1964); A. Lenard, J. Math. Phys. (N. Y.) \textbf{7} 1268 (1966).

\bibitem{Minguzzi} P. Vignolo and A. Minguzzi, Phys. Rev. Lett. \textbf{110}, 020403 (2013).

\bibitem{OHara} K. M. O'Hara, S. L. Hemmer, M. E. Gehm, S. R. Granade, and J. E. Thomas, Science \textbf{298}, 2179 (2002).

\bibitem{Bourdel} T. Bourdel, J. Cubizolles, L. Khaykovich, K. M. F. Magalh\~{a}es, S. J. J. M. F. Kokkelmans, G.V. Shlyapnikov, and C. Salomon, Phys. Rev. Lett. \textbf{91}, 020402 (2003).

\bibitem{Ku} M. J. H. Ku, A. T. Sommer, L.W. Cheuk, and M.W. Zwierlein, Science \textbf{335}, 563 (2012).

\bibitem{TanMom} S. Tan, Ann. Phys. \textbf{323}, 2971 (2008).

\bibitem{Braaten} E. Braaten and L. Platter, Phys. Rev. Lett. \textbf{100}, 205301 (2008).

\bibitem{Stewart} J. T. Stewart, J. P. Gaebler, T. E. Drake, and D. S. Jin, Phys. Rev. Lett. \textbf{104}, 235301 (2010).

\bibitem{TanEnergy} S. Tan, Ann. Phys. \textbf{323}, 2987 (2008).

\bibitem{TanVirial} S. Tan, Ann. Phys. \textbf{323}, 2952 (2008).

\bibitem{SZhang} S. Zhang and A. J. Leggett, Phys. Rev. A \textbf{79}, 023601 (2009).

\bibitem{HaoIJMPB2016} Y. Hao, Y. Song, and X. Fu, arXiv: 1507.03832.

\bibitem{MinguzziPLA} A. Minguzzi, P. Vignolo, and M. Tosi, Phys. Lett. A \textbf{294}, 222 (2002).

\bibitem{Olshanii2003} M. Olshanii and V. Dunjko, Phys. Rev. Lett. \textbf{91}, 090401 (2003).

\bibitem{FourierAnalysis} M. J. Lighthill, Introduction to Fourier Analysis and Generalised Functions (Cambridge University Press, New York, 1958), p. 43.

\end{thebibliography}
\end{document}